\documentclass[aps,prl,10pt,twocolumn,showpacs,superscriptaddress,longbibliography]{revtex4-2}

\usepackage{soul}
\usepackage[colorlinks=true, linkcolor=blue, anchorcolor=blue, citecolor=blue, urlcolor=blue]{hyperref}
\usepackage{amssymb,graphicx,color,amsmath,mathtools,physics}
\usepackage{xfrac,bm,stmaryrd,trimclip,xcolor}

\newcommand{\CGT}{Cr$_2$Ge$_2$Te$_6$}
\newcommand{\CVS}{CsV$_3$Sb$_5$}

\begin{document}
\title{Probing a two-dimensional soft ferromagnet \CGT\ by a tuning fork resonator}
\author{Hengrui Gui}
\affiliation{Center for Correlated Matter and School of Physics, Zhejiang University, Hangzhou 310058, China}
\author{Zekai Shi}
\affiliation{Center for Correlated Matter and School of Physics, Zhejiang University, Hangzhou 310058, China}
\author{Jiawen Zhang}
\affiliation{Center for Correlated Matter and School of Physics, Zhejiang University, Hangzhou 310058, China}
\author{Yu Liu}
\affiliation{Center for Correlated Matter and School of Physics, Zhejiang University, Hangzhou 310058, China}
\author{Huiqiu Yuan}
\email[Corresponding author: ]{hqyuan@zju.edu.cn}
\affiliation{Center for Correlated Matter and School of Physics, Zhejiang University, Hangzhou 310058, China}
\affiliation {Institute for Advanced Study in Physics, Zhejiang University, Hangzhou 310058, China}
\affiliation {Institute of Fundamental and Transdisciplinary Research, Zhejiang University, Hangzhou 310058, China}
\affiliation  {State Key Laboratory of Silicon and Advanced Semiconductor Materials, Zhejiang University, Hangzhou 310058, China}
\author{Lin Jiao}
\email[Corresponding author: ]{lin.jiao@zju.edu.cn}
\affiliation{Center for Correlated Matter and School of Physics, Zhejiang University, Hangzhou 310058, China}

\begin{abstract}
\textbf{Magnetic anisotropy encodes key information about the free-energy landscape of magnetic materials, but its quantitative characterization often requires probes beyond conventional magnetometry. A quartz tuning-fork resonator provides direct access to the magnetotropic susceptibility. Here we use this technique to investigate the magnetic anisotropy of the layered ferromagnet \CGT. The temperature-, field-, and angle-dependent responses are consistently described by a quasi-two-dimensional (2D) easy-axis ferromagnetic model. In particular, the evolution of the magnetotropic susceptibility reveals how the angular profile changes from a conventional cos(2$\theta$) form to a pronounced dip structure as the magnetization approaches directional saturation. These results establish \CGT\ as an ideal reference system for tuning-fork-based magnetotropic measurements. More broadly, they provide a useful framework for distinguishing spin-origin anisotropy from orbital magnetism, as in the case of \CVS~\cite{gui_tf135_2025}. Our work demonstrates that tuning-fork resonators offer a sensitive thermodynamic probe of the rotational stiffness of magnetization in anisotropic low-dimensional magnets.}
\end{abstract}
\maketitle

\section{Introduction}
Magnetic anisotropy is an important, though sometimes underappreciated, property of quantum materials, especially in systems hosting emergent phases such as frustrated magnetism, spin or charge nematicity, and orbital magnetism~\cite{getzlaff_magneticBOOK_2008}. Conventional magnetic probes---such as superconducting quantum interference devices (SQUID) or vibrating sample magnetometer based systems---provide direct access to bulk magnetization, but they are often less suitable for resolving the fine angular dependence of the free energy and the associated rotational stiffness of magnetic order. In this context, magnetotropic susceptibility ($k_{\mathbf n}$), defined as the second angular derivative of the free energy in an applied magnetic field, $k_\textbf{n}(\textbf{H})=\partial^2F(\textbf{H})/\partial\theta^2_\textbf{n}$, has emerged as a particularly useful thermodynamic quantity~\cite{shekhter_magnetotropic_2023}. As emphasized by Shekhter \textit{et al.}, $k_{\mathbf n}$ directly characterizes the curvature of the field-angle-dependent free energy and therefore provides direct access to the rotational stiffness of magnetization in anisotropic systems~\cite{shekhter_magnetotropic_2023}.

Tuning-fork-based mechanical resonators have developed into ultrasensitive probes with broad applications~\cite{A1chen_torque_2018, A2finklerWez_scanning_2012, A3giessibl_Qplus_1998, A4giessibl_qplus_2019, A5giessibl_subatomic_2000, A6gross_chemical_2009, A7seo_MFM_2005, A8vasyukov_SQUIDTF_2013, A9xiang_newSQUIDTF_2023, A10zhou_scanning_2023, A11zhou_imaging_2023}. More recently, they have been used to probe magnetotropic susceptibility in a variety of quantum materials~\cite{modic_resonant_2018, modic_scale-invariant_2021, safari_2024_BaCo2AsO24_TF, zambra_2025_UTe2TF}. By converting the resonant frequency shift into $k_{\mathbf n}$, tuning-fork measurements provide a clean, nearly thermodynamic approach for detecting tiny anisotropies and subtle variations in magnetic response. Owing to their high mechanical quality factor and compatibility with low-temperature, high-field environments, these resonators are particularly powerful in cases where conventional magnetometry lacks sufficient angular sensitivity~\cite{modic_resonant_2018}. For example, tuning-fork measurements on RuCl$_3$ revealed scale-invariant magnetic anisotropy and provided evidence for strong exchange frustration and a possible spin-liquid regime~\cite{modic_scale-invariant_2021}. Likewise, studies of the transverse magnetic response in UTe$_2$ offered important insight into the pairing mechanism of its unconventional superconducting state~\cite{zambra_2025_UTe2TF}. These developments demonstrate the unique advantages of tuning-fork resonators for detecting weak thermodynamic signatures in quantum materials.

Recently, a chiral charge-density-wave (CDW) phase has been proposed in $A$V$_3$Sb$_5$ ($A$ = K, Rb, and Cs), with (controversial) evidence for time-reversal symmetry breaking~\cite{Hasan2021, HuJP_2021_CFphase135, ZhaoZX_2021_uSR135, Guguchia2022, WuL2022, Zhou_2022_135theory,  MollP_2022_135Ctr}. One experimental signature of time-reversal-symmetry breaking is a sharp frequency drop near $H \parallel ab$ measured by a tuning-fork resonator~\cite{gui_tf135_2025}. However, to fully understand the behavior of $k_{\mathbf n}(\theta)$, a systematic calibration against a well-understood magnetic reference system is still needed. A rigorous benchmark based on a well-characterized ferromagnet would not only validate the sensitivity and fidelity of this technique, but also provide a reference framework for interpreting tuning-fork responses in more exotic magnetic states.

\CGT\ has attracted broad experimental and theoretical interest~\cite{li_crxte3theory2_2014, sivadas_magnetictheory1_2015, siberchicot_band_1996, gong_CGTNature_2017, m.mogi_MBECGT_2018, Lin_tricritical_2017,carteaux_1995_basicCGT, huiwenji_CGTsubstrate_2013, zhang_magneticCGTaniso_2016,liu_criticalIsingCGT_2017, WeiLiu_2018_critical2DIsing} as a layered van der Waals ferromagnet with a well-established easy axis along the crystallographic $c$-axis~\cite{carteaux_1995_basicCGT, huiwenji_CGTsubstrate_2013, zhang_magneticCGTaniso_2016, Lin_tricritical_2017}. Some previous studies emphasized its soft-magnetic character and described it within a quasi-2D Heisenberg picture~\cite{gong_CGTNature_2017}. By contrast, other works highlighted the role of single-ion anisotropy and anisotropic intralayer exchange, which endow the system with pronounced easy-axis, Ising-like behavior rather than that of an isotropic Heisenberg magnet~\cite{zhuang_computationalIsingSpin_2015, liu_criticalIsingCGT_2017, huiwenji_CGTsubstrate_2013, zhang_magneticCGTaniso_2016, WeiLiu_2018_critical2DIsing, Yuliu_CGTentropy_2019}. In the present work, our goal is not to settle the universality-class assignment of \CGT, but to exploit the fact that it combines extremely soft ferromagnetism with robust out-of-plane easy-axis anisotropy. Establishing this evolution experimentally is valuable not only for understanding the magnetotropic susceptibility of a soft easy-axis ferromagnet, but also for providing a reference framework for interpreting more unconventional tuning-fork responses, including those associated with orbital magnetism in \CVS~\cite{gui_tf135_2025}. 
For this purpose, here we use a quartz tuning-fork resonator to investigate the temperature-, field-, and angle-dependent magnetotropic response of \CGT. We show that the observed evolution of the resonant frequency shift is consistently captured by an easy-axis ferromagnetic model with quasi-2D character, thereby establishing \CGT\ as a benchmark system for interpreting tuning-fork responses in more unconventional magnetic states.

\section{Results}

\begin{figure}
			\centering
			\includegraphics[width=1.0\linewidth]{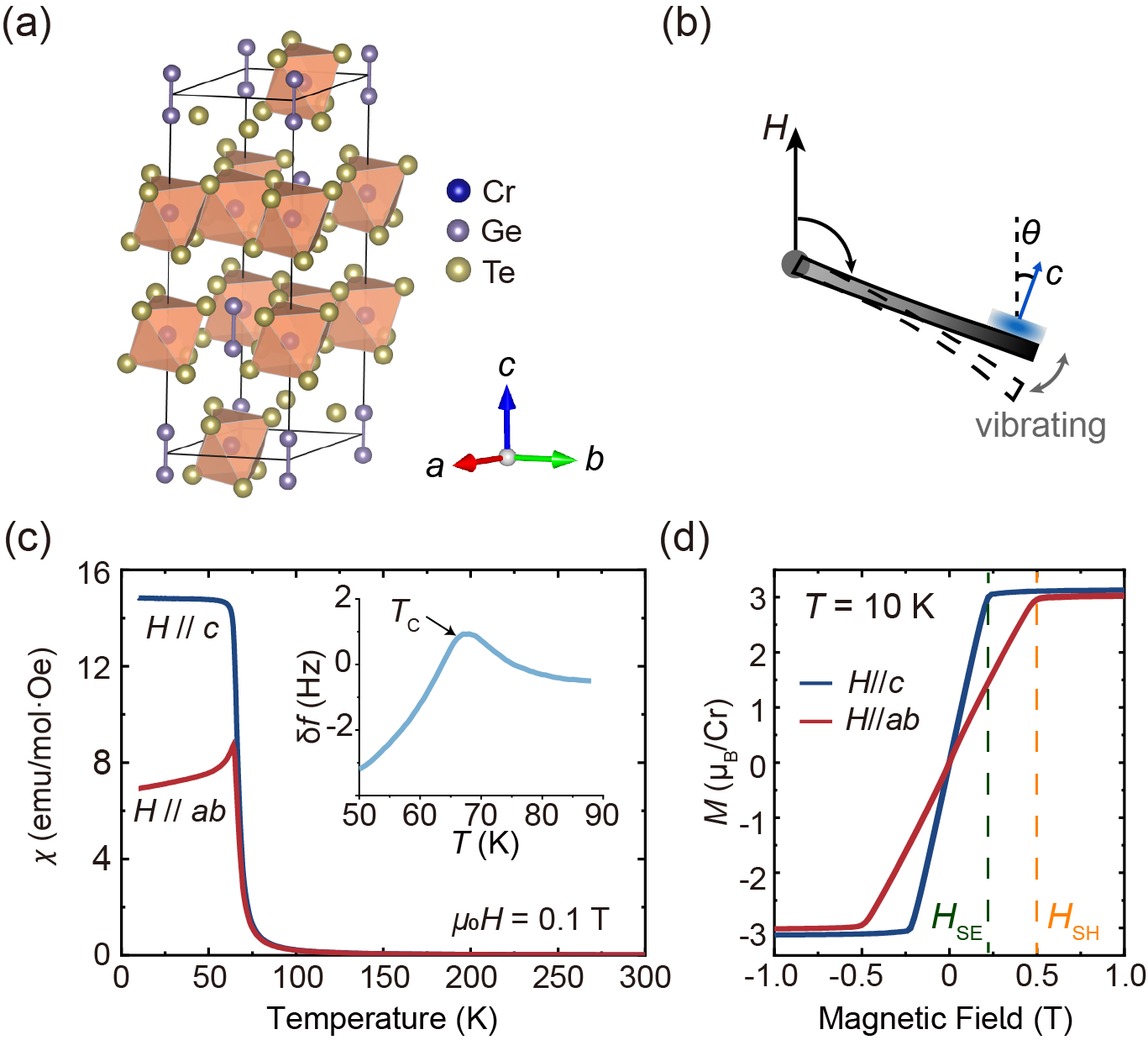}
			\caption{(a) Crystal structure of \CGT. (b) Schematic illustration of the tuning-fork measurement setup. The sample is mounted at the front end of the cantilever, with the rotation axis perpendicular to the $c$-axis. $\theta$ is defined as the angle between the magnetic field $H$ and the $c$-axis. (c) Temperature dependence of the magnetic susceptibility of \CGT\ for $H \parallel ab$ and $H \parallel c$. The inset shows the temperature dependence of the resonant frequency shift, $\delta f=f(0.5~\mathrm{T})-f(0~\mathrm{T})$, for $H \parallel ab$. (d) Magnetization curves measured at 10~K, showing the saturation fields for $H \parallel ab$ and $H \parallel c$.}
        
			\label{fig1}
\end{figure}

Figure~\ref{fig1}(a) shows the crystal structure of bulk \CGT. Each unit cell consists of three van der Waals layers stacked in an ABC sequence along the $c$-axis. The Cr atoms occupy the centers of slightly distorted Te octahedra and form a honeycomb lattice within each layer, while the Te atoms act as interlayer spacers. The Ge atoms form Ge$_2$Te$_6$ ethane-like units because of the short Ge--Ge bond.
Previous studies established that the magnetic easy axis of \CGT\ lies along the crystallographic $c$-axis, with negligible in-plane anisotropy~\cite{WeiLiu_2018_critical2DIsing}. To probe this anisotropy thermodynamically, we measured the magnetotropic susceptibility using a tuning-fork resonator. As illustrated in Fig.~\ref{fig1}(b), the sample was mounted on the tip of an Akiyama probe and rotated from $H \parallel c$ toward $H \parallel ab$~\cite{Akiyamaprobe}. The basic magnetic properties of \CGT\ were independently characterized by SQUID magnetometry, as shown in Figs.~\ref{fig1}(c) and \ref{fig1}(d). The Curie temperature, $T_{\mathrm C}=61$~K, is consistently identified by both SQUID and tuning-fork measurements [Fig.~\ref{fig1}(c)], in agreement with previous reports~\cite{carteaux_1995_basicCGT, zhang_magneticCGTaniso_2016}. In addition, the $M(H)$ curves saturate at about $\mu_0H_{\mathrm{SE}} \sim 0.22$~T for the easy axis ($H \parallel c$) and $\mu_0H_{\mathrm{SH}} \sim 0.5$~T for the hard direction ($H \parallel ab$). No visible hysteresis is detected down to 10~K (see Fig.~S1), consistent with the extremely soft ferromagnetic character reported previously~\cite{gong_CGTNature_2017}. More importantly, the saturated magnetization reaches nearly the same value for both field orientations above $\sim 0.5$~T, indicating that the response becomes nearly isotropic once the Zeeman energy dominates over the intrinsic easy-axis anisotropy.

To place the tuning-fork results in context, we briefly summarize the expected behavior of the magnetotropic susceptibility $k_{\mathbf n}$ (proportional to the resonant frequency shift $\Delta f$) for several representative magnetic responses~\cite{shekhter_magnetotropic_2023}. For an anisotropic paramagnet with a linear but direction-dependent $M(H)$ curve, the angular dependence of $k_{\mathbf n}$ follows $(\chi_c-\chi_{ab})$$H^2$cos(2$\theta$) behavior, where $\theta$ is the angle between $H$ and $c$-axis, the rotation axis is the $b$-axis, and $\chi_c$ ($\chi_{ab}$)  is the magnetic susceptibility for $H \parallel c$ ($H \parallel ab$). In this limit, the rotational amplitude is directly related to the magnetic anisotropy through $\Delta k_{\mathbf n} = |k_{\mathbf n}(90^\circ) - k_{\mathbf n}(0^\circ)| \propto (\chi_c-\chi_{ab})$.
By contrast, for an isolated $S=1/2$ moment with an anisotropic $g$-factor tensor, the free energy takes the form $F(H)$ = $-T$ln(2cosh($\sqrt{a}/2T))$, with $a=\mu^2_B (g^2_{aa}B^2_{a}+g^2_{bb}B^2_{b}+g^2_{cc}B^2_{c})$, and the $g$-factor tensor is diagonal in $a, b, c$ basis~\cite{shekhter_magnetotropic_2023}. In this scenario, $k_{\mathbf n}(\theta)$ generally deviates from a simple cos(2$\theta$) form.
For a conventional easy-axis ferromagnet, the $M(H)$ curve can be approximated by a $\tanh(H)$-like dependence~\cite{feynman1964}. When the sample is rotated from the easy axis toward the hard axis, the magnetic torque ${\mathbf \tau}(\theta)$ and magnetotropic susceptibility $k_{\mathbf n}(\theta)$ can be written as: 
\begin{equation}
    \mathbf{\tau}(\theta) = \mu_0\textbf{n} \cdot \textbf{M} \times \textbf{H} \propto h*\mathrm{tanh}(h\mathrm{cos}(\theta))\mathrm{sin}(\theta),
\end{equation}
\begin{equation}
\begin{split}\label{eq2}
        k_{\mathbf n}(\theta) =& \frac{d\tau_\textbf{n}(\textbf{H})}{d\theta_\textbf{n}} \propto -h^2\mathrm{sech}(h\mathrm{cos}(\theta))^2\mathrm{sin}(\theta)^2\\
    &+h\mathrm{cos}(\theta)\mathrm{tanh}(h\mathrm{cos}(\theta)),
\end{split}
\end{equation}
where $h=H/H_{\mathrm S}$ is the magnetic field normalized by the saturation field. This form naturally gives rise to a sharp dip in $k_{\mathbf n}(\theta)$ when the field approaches the hard-axis direction. Notably, the same phenomenological form was recently used to describe the loop-current phase in \CVS, which exhibits a strong ferromagnet-like anisotropic response~\cite{gui_tf135_2025}.

\begin{figure}{
			\centering
			\includegraphics[width=1.0\linewidth]{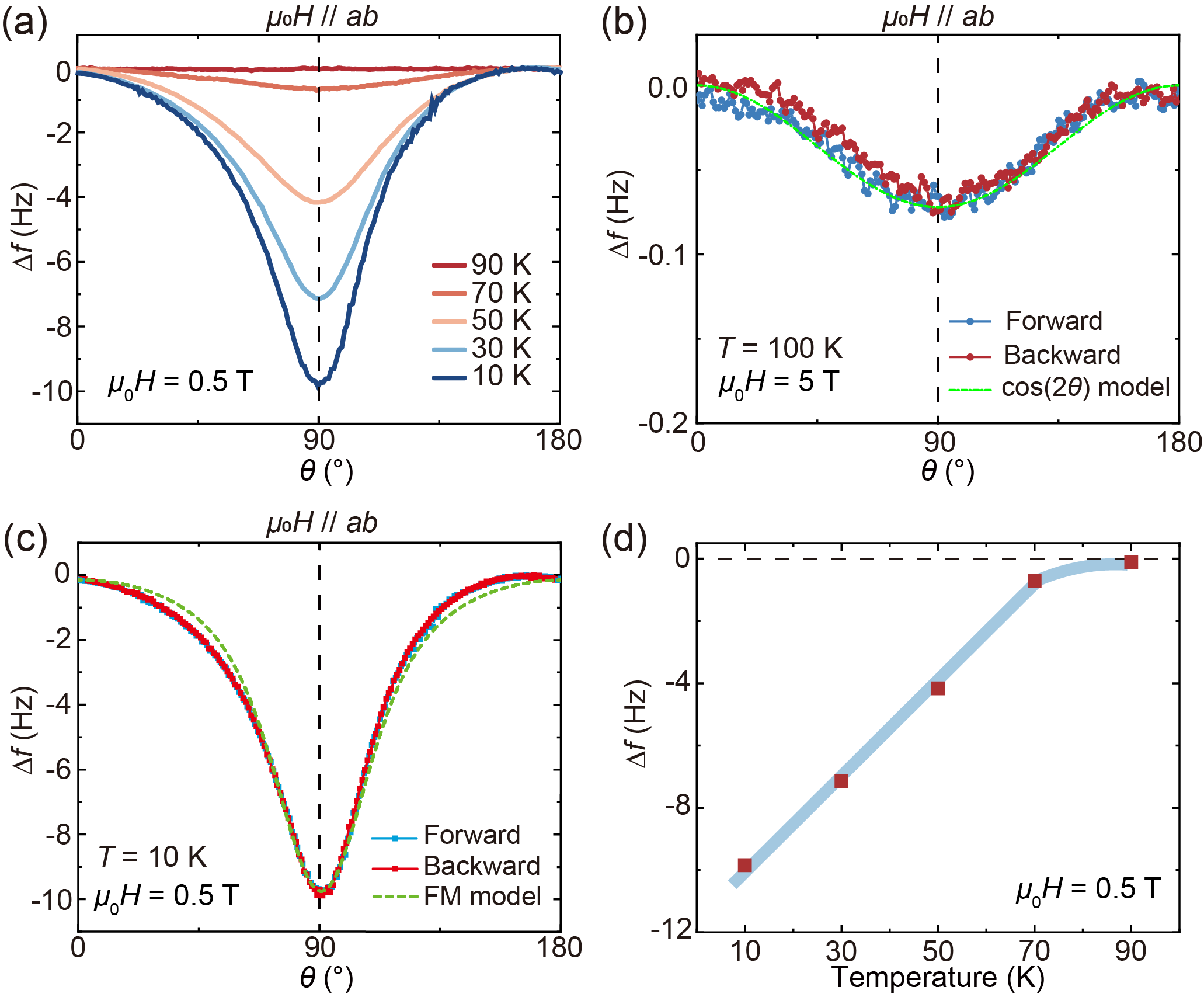}
\caption{(a) Angular dependence of $\Delta f$ at different temperatures for $\mu_0H=0.5$~T. (b) Forward and backward rotation measurements of $\Delta f$ at 100~K and 5~T in the paramagnetic phase. The green dashed line shows a $\cos(2\theta)$ fit. (c) Forward and backward rotation measurements of $\Delta f$ at 10~K and 0.5~T in the ferromagnetic phase. The green dashed line shows the fit to the phenomenological easy-axis ferromagnetic model. (d) Temperature dependence of $\Delta f$ at $\theta=90^\circ$, extracted from panel (a).}
\label{fig2}}
\end{figure}

Figures~\ref{fig2}(a-c) show the angular dependence of the resonant frequency shift at different temperatures. Here, $\Delta f(\theta)$ is defined as $\Delta f(\theta)=f(\theta)-f(0^\circ)$. Above $T_{\mathrm C}$, $\Delta f(\theta)$ remains very small at low fields. At a sufficiently large field (for example, 5~T), however, a simple $\cos(2\theta)$ dependence becomes clearly resolved, as expected for an anisotropic paramagnetic state (see Fig.~\ref{fig2}(b)). Upon cooling toward $T_{\mathrm C}$, however, \CGT\ develops a much stronger anisotropic response: at 0.5~T, $\Delta f(\theta)$ progressively deviates from the $\cos(2\theta)$ form and evolves into a profile with a pronounced dip at $\theta=90^\circ$. The depth of this dip increases steadily with decreasing temperature, as summarized in Figs.~\ref{fig2}(a) and \ref{fig2}(d). Figure~\ref{fig2}(d) further shows that the amplitude of $\Delta f$ continues to increase down to the lowest measured temperature, even though $\chi_c(T)$ nearly saturates below $T_{\mathrm C}$. This behavior originates from the clear suppression of $\chi_{ab}(T)$ upon cooling (see Fig.~\ref{fig1}(c)), which enhances the magnetic anisotropy sensed by the tuning-fork resonator. Therefore, the tuning-fork measurement is particularly sensitive to weak anisotropic contributions that are much less evident in the bulk susceptibility itself.

Most importantly, below $T_{\mathrm C}$ the angular dependence of $\Delta f(\theta)$ is well described by the phenomenological easy-axis ferromagnetic model introduced in Eq.~\ref{eq2}, as shown in Fig.~\ref{fig2}(c). The resulting line shape, especially the sharp curvature near $\theta=90^\circ$, is reminiscent of the low-temperature response observed in \CVS, where it was attributed to orbital magnetism with strong directional locking~\cite{gui_tf135_2025}. At the same time, an important difference is evident: forward and backward rotations in \CGT\ produce nearly identical traces without measurable hysteresis, consistent with its extremely soft ferromagnetic character. This contrasts with \CVS, where the response exhibits much slower dynamics~\cite{gui_tf135_2025}.
Another notable feature is that the sharp dip develops specifically near $H \parallel ab$ ($\theta=90^\circ$). During the rotation, the field component along the easy axis remains sufficient to polarize the magnetization once it exceeds the easy-axis saturation scale $H_{\mathrm{SE}}$. As a result, the most abrupt change in $k_{\mathbf n}$ occurs when the field approaches the hard-axis direction, giving rise to the pronounced dip structure in $\Delta f(\theta)$.

\begin{figure}
			\centering
			\includegraphics[width=1.0\linewidth]{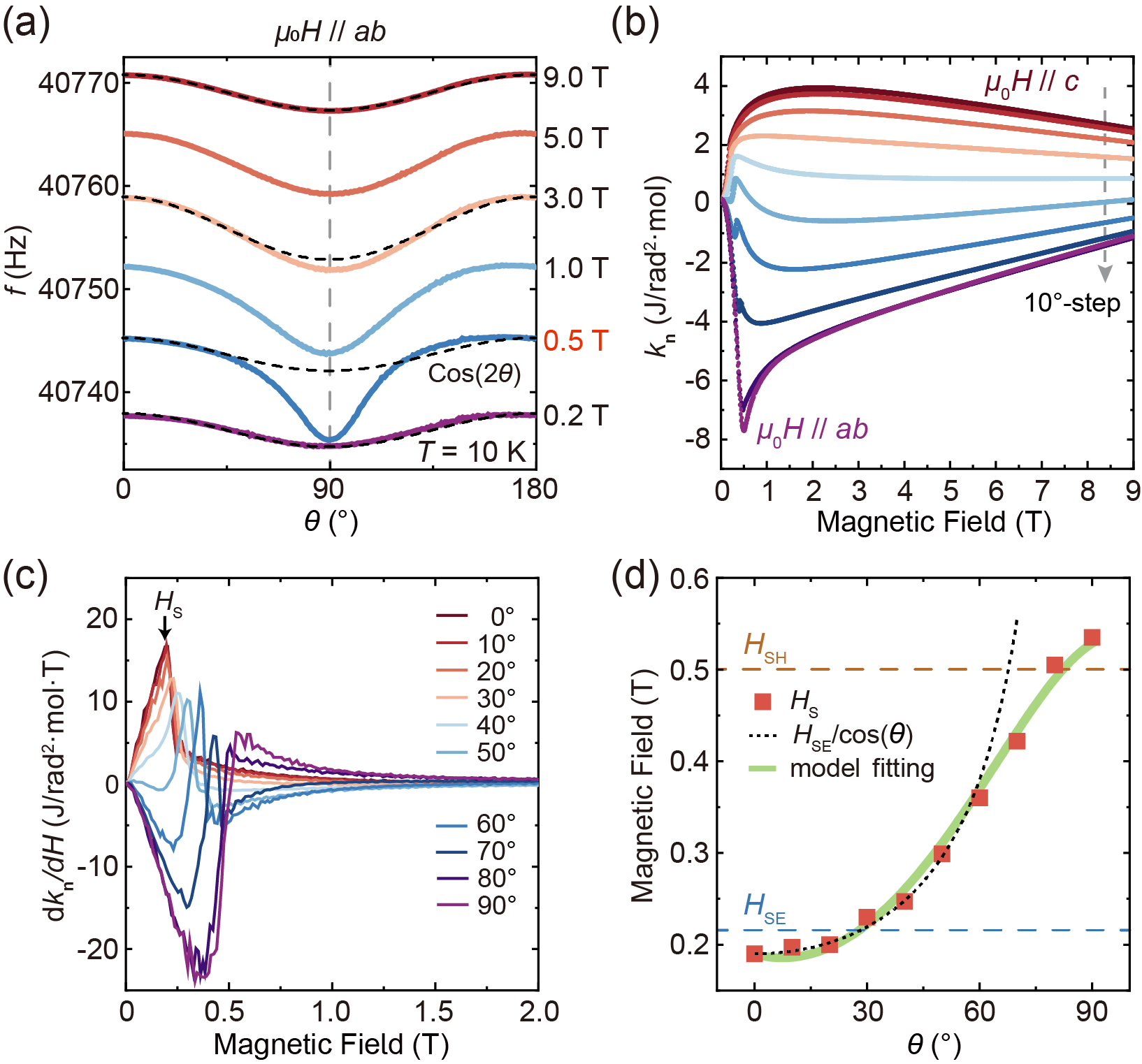}
            \caption{(a) Angular dependence of the resonant frequency at different magnetic fields at 10~K. (b) Field dependence of $k_{\mathbf n}$ at selected field orientations. (c) Derivative of $k_{\mathbf n}$ with respect to $H$. The peak values are used to define the saturation field $H_{\mathrm S}$. (d) Angular dependence of $H_{\mathrm S}(\theta)$ extracted from panel (c), together with two different fits. The black dashed line is proportional to $1/\cos\theta$. The green solid line is obtained from the energy minimization based on Eqs.~(\ref{eq3}) and (\ref{eq4}).}
			
			\label{fig3}
\end{figure}

To track the field evolution of the magnetic anisotropy, we measured $f(\theta)$ at different magnetic fields, as shown in Fig.~\ref{fig3}(a). In the low-field regime below $H_{\mathrm{SE}}$, the $M(H)$ curves remain in the linear-response regime for all field orientations, and $f(\theta)$ is therefore well described by a simple $\cos(2\theta)$ dependence. Once the field reaches $H_{\mathrm{SE}}$, however, a pronounced dip develops, as already seen in Fig.~\ref{fig2}(c). In Fig.~\ref{fig3}(a), this strong deviation from the $\cos(2\theta)$ form becomes evident around 0.5~T. At still higher fields, however, the angular dependence evolves back toward a nearly $\cos(2\theta)$ profile. In particular, for $H \gg H_{\mathrm{SH}}$ (for example, above 3~T), the difference between the $M(H)$ curves for $H \parallel c$ and $H \parallel ab$ becomes very small, and the response approaches a nearly isotropic limit. This behavior indicates that, in sufficiently strong fields, Zeeman polarization overwhelms the intrinsic magnetocrystalline anisotropy and drives the magnetization toward a fully saturated state~\cite{Selter_2020_mboundary}. This nearly isotropic state at high fields resembles the field-polarized paramagnetic state found in 2D antiferromagnet material CrPS$_4$ by a similar technique~\cite{cpl_ZJL_2024}.

To further examine this crossover, we measured $f(H,\theta)$ and converted the frequency shift into $k_{\mathbf n}(H,\theta)$, as shown in Fig.~\ref{fig3}(b) (the conversion procedure is described in the Supplementary Information). As the system approaches the fully polarized state, $k_{\mathbf n}(H,\theta)$ gradually tends toward zero, consistent with the loss of angular curvature in the free energy. This trend is clearly seen in Fig.~\ref{fig3}(b). An important consequence of the $k_{\mathbf n}(H,\theta)$ data is that the magnetic moment can be estimated directly from the anisotropic response. For a strongly 2D magnetic system, the $c$-axis moment can be approximated as $m=\Delta k_{\mathbf n}(H)/\mu_0H$~\cite{gui_tf135_2025}. Using $\Delta k_{\mathbf n} \approx 11$~J/mol/rad$^2$ at 0.5~T, where $\Delta k_{\mathbf n}$ reaches its maximum, we obtain an effective moment of $\sim 2\mu_{\mathrm B}$/Cr. Although this value is somewhat smaller than the measured moment of Cr$^{3+}$ in Fig.~\ref{fig1}(d), it remains reasonably close in magnitude, indicating that this approximation still provides a useful estimate of the magnetic moment despite the finite in-plane contribution.

Another important feature is the kink in $k_{\mathbf n}(H,\theta)$, which becomes more evident in the derivative plot shown in Fig.~\ref{fig3}(c). These kinks appear near the saturation field $H_{\mathrm S}(\theta)$. Figure~\ref{fig3}(d) summarizes the angular evolution of $H_{\mathrm S}(\theta)$ extracted from the kink positions. For the limiting cases of $H \parallel c$ and $H \parallel ab$, the inferred saturation fields agree well with the values obtained independently from magnetization measurements (dashed lines). Between $0^\circ$ and $90^\circ$, $H_{\mathrm S}$ evolves smoothly with field orientation. Previous electron-spin-resonance, ferromagnetic-resonance, and magnetization studies also reported signatures of an angle-dependent saturation field in \CGT~\cite{Zeisner_2019_CGTangle, WeiLiu_2018_critical2DIsing}. To understand the behavior of $H_{\mathrm S}(\theta)$ in Fig.~\ref{fig3}(d), we first compare the data with a simple $1/\cos(\theta)$ form, which takes into account only the field component along the easy axis and neglects the in-plane magnetization. This approximation describes the data reasonably well up to about $60^\circ$. Beyond this angle, however, the in-plane field component can no longer be ignored, and $H_{\mathrm S}(\theta)$ gradually bends toward the hard-axis saturation field $H_{\mathrm{SH}}$. To capture this curvature more realistically, we consider the competition between magnetocrystalline anisotropy and Zeeman coupling. The corresponding magnetic free-energy density can be written as
\begin{equation}
F_{\mathrm{mag}} = F_Z + F_c = -\mu_0 MH\mathrm{cos}(\theta-\phi)+K_1\mathrm{cos}^2(\phi),
\label{eq3}
\end{equation}
where $\phi$ represents the angle of the magnetic moment measured from the $c$-axis. The equilibrium condition is obtained by minimizing $F_{\mathrm{mag}}$ with respect to $\phi$:~\cite{getzlaff_magneticBOOK_2008}:
\begin{equation}
2K_1\mathrm{cos}(\phi)\mathrm{sin}(\phi)+\mu_0MH\mathrm{sin}(\theta-\phi)=0.
\label{eq4}
\end{equation}
Using $K_1$ = $\frac{1}{2}H_{\mathrm S}M_{\mathrm S}$ = 3.5$\times10^5$~erg/cm$^3$~\cite{cullity_2009_introduction, zhang_magneticCGTaniso_2016}, we obtain a significantly improved description of $H_{\mathrm S}(\theta)$, shown by the green curve in Fig.~\ref{fig3}(d). This analysis further highlights the strength of magnetotropic susceptibility as a quantitative probe of anisotropic magnetic response.

\begin{figure}
			\centering
			\includegraphics[width=1.0\linewidth]{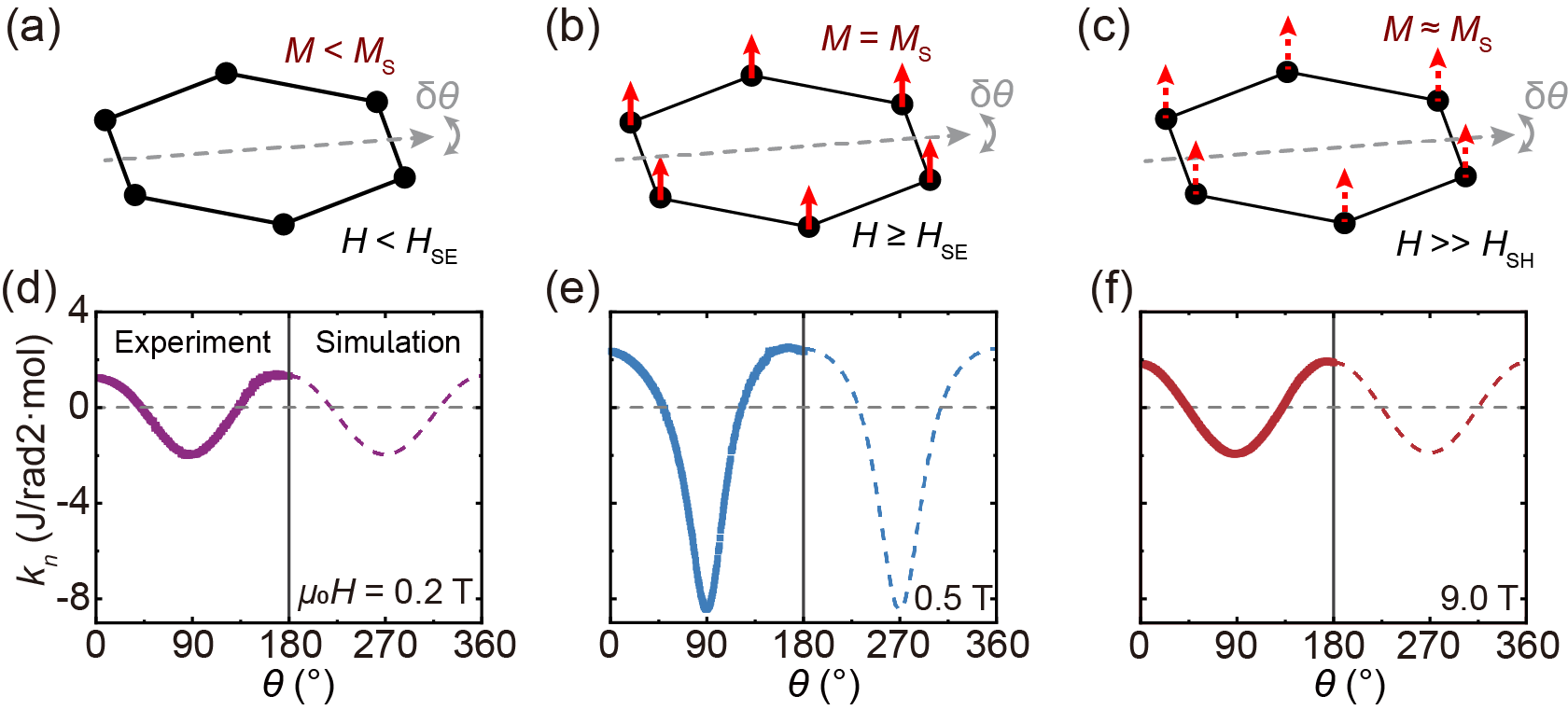}
			\caption{Schematic illustration of the magnetic response of Cr moments in \CGT\ under a rotating magnetic field at low temperature. Panels (a)-(c) show the spin configurations in three field regimes: (a) $H<H_{\mathrm{SE}}$, (b) $H_{\mathrm{SE}} \leq H < H_{\mathrm{SH}}$, and (c) $H \gg H_{\mathrm{SH}}$. Panels (d)-(f) show the corresponding measured (left panel) and simulated (right panel) angular dependence of the magnetotropic susceptibility.}
			\label{fig4}
\end{figure}

\section{Discussions and Conclusion}

Our results provide a precise temperature-, field-, and angle-resolved characterization of the easy-axis ferromagnetic response in \CGT. 
By exploiting its pronounced magnetic anisotropy together with its extremely soft ferromagnetic character, we establish \CGT\ as an ideal benchmark system for calibrating tuning-fork-based magnetotropic measurements. 
Figure~\ref{fig4} summarizes the measured angular dependence of $k_{\mathbf n}(\theta)$ (equivalently, $\Delta f(\theta)$) at representative magnetic fields below $T_{\mathrm C}$, together with simulated curves and schematic spin configurations. 
At low fields, the magnetization remains in the linear-response regime and the angular dependence follows an approximate $\cos(2\theta)$ form. 
As the field increases, the $c$-axis magnetization gradually approaches saturation, which deforms the original $\cos(2\theta)$ profile and produces a pronounced dip near the hard-axis direction. 
At sufficiently high fields ($H \gg H_{\mathrm{SH}}$), the net magnetic moment again follows the external field for nearly all orientations, and the angular response evolves back toward a simple $\cos(2\theta)$ form.

The systematic evolution of $k_{\mathbf n}(H,\theta)$ not only characterizes the easy-axis ferromagnetism of \CGT, but also offers important insight into our recent observations in \CVS~\cite{gui_tf135_2025}. 
In \CVS, a qualitatively similar evolution of $\Delta f(\theta)$ is observed. 
Above the CDW transition temperature $T_{\mathrm{CDW}}$, $\Delta f(\theta)$ is nearly featureless. 
The opening of the CDW gap introduces band anisotropy, giving rise to a conventional $\cos(2\theta)$ angular dependence. 
Upon further cooling below $\sim 30$~K, a loop-current CDW phase is stabilized, which generates an orbital magnetization rigidly constrained along the $c$ axis. 
As a result, a strong Ising-like anisotropy develops, and its fingerprint in $\Delta f(\theta)$ is captured by the tuning-fork resonator in close analogy to the behavior observed in \CGT. 
In particular, at low temperature $\Delta f(\theta)$ in \CVS\ also evolves from a $\cos(2\theta)$ line shape to a sharp dip at $90^\circ$ once the magnetic field reaches about 0.15~T.

Despite these similarities, a crucial distinction emerges in the high-field regime. 
In the loop-current phase of \CVS, the amplitude of $\Delta f(\theta)$ increases monotonically with magnetic field up to the highest measured field of 15~T (see the peer-review file in Ref.~\cite{gui_tf135_2025}). 
This behavior is in sharp contrast to \CGT, where the magnetic anisotropy is progressively suppressed once the magnetization saturates along different field directions, and $f(\theta)$ eventually recovers a simple $\cos(2\theta)$ form. 
In \CGT, this recovery reflects the fact that a sufficiently strong magnetic field overwhelms the intrinsic magnetocrystalline anisotropy, thereby driving the system toward a nearly isotropic polarized state. 
In \CVS, by contrast, the loop-current-induced orbital moment is rigidly locked to the $c$-axis and cannot be rotated by an external field~\cite{gui_tf135_2025}. 
Consequently, the anomalous behavior of $\Delta f$ persists to high fields, and neither a strong suppression of $\Delta f$ nor a restoration of the $\cos(2\theta)$ behavior is expected. 
This comparison demonstrates that the magnetic anisotropy detected in \CVS\ is ferromagnet-like in phenomenology, yet fundamentally distinct in origin, reflecting orbital rather than spin magnetism. The contrast between these two systems shows that the low-field angular line shape alone is insufficient to identify the microscopic origin of the anisotropy; the high-field evolution of $k_{\mathbf n}$ provides the decisive diagnostic.

In conclusion, by taking \CGT\ as a reference system, we identify universal features of tuning-fork-based magnetotropic measurements across paramagnetic and ferromagnetic regimes. 
This technique provides a systematic way to study the temperature-, field-, and angle-dependent magnetotropic susceptibility. 
From these measurements, one can not only track phase transitions, but also extract the magnetic moment and infer changes in magnetic symmetry. 
Most importantly, the magnetic-field dependence of $k_n$ provides a practical experimental criterion for distinguishing spin magnetism from (molecular) orbital magnetism through their fundamentally different high-field evolution. 
As a minimally invasive and highly sensitive thermodynamic probe, the tuning-fork resonator technique is therefore poised to play a distinctive role in the study of correlated quantum materials.

\section*{acknowledgments}
This work was supported by the National Key R\&D Program of China (Grant No. 2022YFA1402200, 2023YFA1406100, 2025YFA1411501), the National Natural Science Foundation of China (Grants Nos. 12374151, 12550401, 12350710785, W2511006), and the Zhejiang Provincial Natural Science Foundation of China (Grant No. LR25A040003).

\bibliographystyle{apsrev4-2}

%

\end{document}